\documentclass{article}
\usepackage{spconf, amsmath, amssymb, graphicx}

\usepackage{booktabs}
\usepackage[acronym, nomain, nopostdot]{glossaries}
\usepackage{color}
\usepackage{import}
\usepackage{gensymb}
\usepackage{nohyperref}  
\usepackage{url}  

\usepackage{todonotes}

\newacronym{ASR}{ASR}{automatic speech recognition}
\newacronym{AV}{AV}{audiovisual}
\newacronym{BKF}{BKF}{backprop Kalman filter}
\newacronym{CNN}{CNN}{convolutional neural network}
\newacronym{DoA}{DoA}{direction-of-arrival}
\newacronym{DSW}{DSW}{dynamic stream weight}
\newacronym{EM}{EM}{expectation maximization}
\newacronym{FFT}{FFT}{fast Fourier transform}
\newacronym{FP}{FP}{false positive}
\newacronym{FR}{FR}{frame recall}
\newacronym{GRU}{GRU}{gated recurrent unit}
\newacronym{KF}{KF}{Kalman filter}
\newacronym{LDS}{LDS}{linear dynamical system}
\newacronym{LE}{LE}{localization error}
\newacronym{LSTM}{LSTM}{long short-term memory}
\newacronym{MSE}{MSE}{mean squared error}
\newacronym{PDF}{PDF}{probability density function}
\newacronym{ReLU}{ReLU}{rectified linear unit}
\newacronym{RMSE}{RMSE}{root mean squared error}
\newacronym{SNR}{SNR}{signal-to-noise ratio}
\newacronym{SRP-PHAT}{SRP-PHAT}{steered response power phase transform}
\newacronym{STFT}{STFT}{short-time Fourier transform}
\newacronym{YOLO}{YOLO}{\emph{you only look once}}

\title{DATA FUSION FOR AUDIOVISUAL SPEAKER LOCALIZATION:\\EXTENDING DYNAMIC STREAM WEIGHTS TO THE SPATIAL DOMAIN}

\begin{document}
\ninept
\name{\parbox{\linewidth}{\centering
		Julio Wissing$^{\star}$ \quad
		Benedikt Boenninghoff$^{\star}$ \quad
		Dorothea Kolossa$^{\star}$ \quad
		Tsubasa Ochiai$^{\dagger}$ \quad 
		Marc Delcroix$^{\dagger}$ \quad 
		Keisuke Kinoshita$^{\dagger}$ \quad 
		Tomohiro Nakatani$^{\dagger}$ \quad 
		Shoko Araki$^{\dagger}$ \quad
		Christopher Schymura$^{\star}$}}
\address{$^{\star}$Institute of Communication Acoustics, Ruhr University Bochum, Bochum, Germany \\ $^{\dagger}$NTT Communication Science Laboratories, NTT Corporation, Kyoto, Japan }
\maketitle
\begin{abstract}
Estimating the positions of multiple speakers can be helpful for tasks like automatic speech recognition or speaker diarization. Both applications benefit from a known speaker position when, for instance, applying beamforming or assigning unique speaker identities. Recently, several approaches utilizing acoustic signals augmented with visual data have been proposed for this task. However, both the acoustic and the visual modality may be corrupted in specific spatial regions, for instance due to poor lighting conditions or to the presence of background noise. This paper proposes a novel audiovisual data fusion framework for speaker localization by assigning individual dynamic stream weights to specific regions in the localization space. This fusion is achieved via a neural network, which combines the predictions of individual audio and video trackers based on their time- and location-dependent reliability. A performance evaluation using audiovisual recordings yields promising results, with the proposed fusion approach outperforming all baseline models. 
\end{abstract}
\section{Introduction}
\label{sec:introduction}
Data fusion is a process aimed at combining several data sources to accomplish a task more reliably than it would be possible with just one individual sensor. This has several useful applications, ranging from medical diagnostics~\cite{Mendes2016} to robotics~\cite{Kam1997} and smart buildings~\cite{Li2015}. Tracking one or more speakers by means of acoustic and visual sensors also relies on an appropriate fusion of sensory inputs. This process is performed unconsciously in human listening, since humans naturally combine visual and acoustic stimuli in estimating the position of a person speaking or a sound source emitting noise~\cite{Hershey2000}.

Recent works in the domain of audiovisual speaker localization showed promising results utilizing different data fusion strategies. For example, \cite{Ban2018, Alameda-Pineda2019} introduce a variational Bayesian approximation to optimally merge acoustic and visual data for combined localization and tracking. A related approach introduced in~\cite{Gebru2014} uses an \gls{EM} algorithm for weighted clustering in the audiovisual observation space. Other probabilistic techniques for audiovisual speaker tracking based on particle filters were proposed in~\cite{GaticaPerez2003, GaticaPerez2007}. Additionally, the work in~\cite{Nakadai2002} combined audiovisual speaker localization with speech separation on a robotic platform in a real-time processing system. A similarly designed audiovisual fusion system for smart rooms \cite{Busso2005} allows for joint localization and identification of persons present in a room.
\begin{figure}[t]
	\centering
	\includegraphics[width=0.8\linewidth]{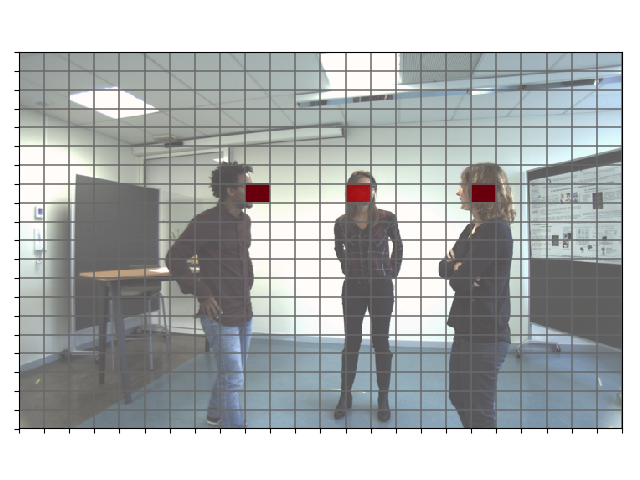}
    \vspace*{-0.7cm}
	\caption{Frame from the AVDIAR dataset~\cite{Gebru2018}, overlayed with speaker positions predicted by the proposed fusion model. Note that the model utilizes a discrete grid to represent the localization space, where each grid-cell is associated with a speaker presence probability, described in Sec.~\ref{subsec:data_fusion}.}
	\vspace*{-0.3cm}
	\label{fig:fusion_tracking_result}
\end{figure}

A classical approach to combine different sensors is the \gls{KF}~\cite{kalman1960}, where inference in a \gls{LDS} is carried out to estimate all latent variables of interest; such as positions, velocities or accelerations, based on noisy measurement data. Extending this concept to \glspl{DSW} allows us to weight the contributions of each input based on their instantaneous reliability. This concept was originally proposed in the context of audiovisual \gls{ASR} \cite{Meutzner2017}, and has proven valuable for speaker identification \cite{Schoenherr2016}, but was also recently adopted for speaker localization and tracking~\cite{Schymura2019, Schymura2020}. 

All of these approaches specify \glspl{DSW} as time-dependent weighting factors for the acoustic and visual modalities. In this study, we introduce a fundamental update: Since the reliability of the acoustic and visual streams may also vary over time and \emph{location}, we extend the idea of \glspl{DSW} to the spatial dimension, as depicted in Fig.~\ref{fig:fusion_tracking_result}. Therefore, we introduce a spatial weighting matrix to replace the scalar \glspl{DSW} used in previous works. Compared to previously proposed approaches, spatial \glspl{DSW} can effectively capture specific position-dependent sensor reliability properties, e.g., bright lighting conditions coming from a certain direction or directional acoustic noise sources. Whereas scalar \glspl{DSW} can only provide a global estimate of sensor reliability, the model proposed here provides a more fine-grained location-dependent framework for data fusion.

\section{System description}
\label{sec:system_description}
\begin{figure}[t]
	\centering
	\includegraphics[width=0.8\linewidth]{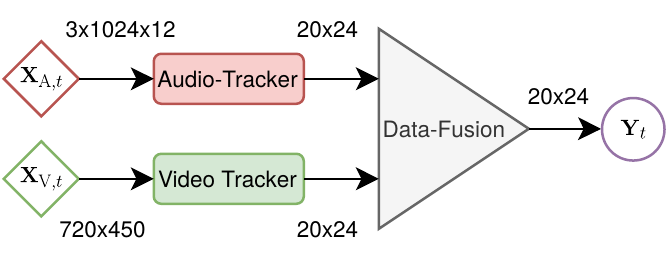}
		\vspace*{-0.1cm}
	\caption{High-level overview of the proposed data fusion framework. The acoustic and visual inputs \(\boldsymbol{\mathrm{X}}_{\text{A}, t}\) and \(\boldsymbol{\mathrm{X}}_{\text{V}, t}\) are first processed via unimodal trackers and then combined in a dedicated fusion network, producing a grid of speaker presence probabilities  \(\boldsymbol{\mathrm{Y}}_{t}\) as output.}
		\vspace*{-0.2cm}
	\label{fig:overview}
\end{figure}

The proposed data fusion network consists of two building-blocks: the individual unimodal tracking networks and the \gls{DSW}-based data fusion network, as depicted in Fig~\ref{fig:overview}.

\subsection{Audio tracker}
\label{subsec:audio_tracker}
We adopt the unimodal audio tracker from the DOAnet architecture~\cite{Adavanne2018}. Here, a multi-channel \gls{STFT} is performed on the individual microphone signals. Time-frequency analysis is conducted on the 6-channel acoustic inputs using a window length of 1920 samples with no overlap and a \gls{FFT} size of 2048. This results in a frame length of \(40\,\text{ms}\), matching the length of one video frame in the utilized dataset. A context of one frame in both temporal directions around the \(t\)-th time step is used, which results in a 3-dimensional acoustic input tensor \(\boldsymbol{\mathrm{X}}_{\text{A}, t}\) of dimensions \(3 \times 1024 \times 12\), representing the time, frequency and channel dimensions. The third dimension is spanned by the stacked amplitude and phase of all 6 audio channels.

Similar to the DOAnet architecture, the acoustic input tensor is fed through 4 convolutional stages first. Herein, each stage is composed of a 2-dimensional convolution with \(64\) filters and a kernel size of \(3 \times 3\), batch normalization, \gls{ReLU} nonlinearity, max-pooling of size \(1 \times 8\), and a dropout layer with a rate of \(0.5\). Subsequently, the tensor is flattened, yielding a dimensionality of \(3 \times 128\), and input to a bidirectional \gls{GRU} with two layers. The output of these recurrent layers is then fed to a classification network, composed of three fully connected layers with \(2048\), \(1024\) and \(512\) neurons and \glspl{ReLU}, respectively. Lastly, the output layer is constructed using another fully connected layer with \(480\) neurons and sigmoid activations. The outputs are reshaped to a grid with \(20 \times 24\) cells, as depicted in Fig.~\ref{fig:fusion_tracking_result}. The grid size was chosen to roughly represent the size of a face in the utilized dataset. Using a binary indicator variable \(\zeta_{i, j}\) corresponding to the \(i\)-th row and \(j\)-th column of the grid, the probability that a speaker is located at this particular grid point given only the current acoustic observation can be expressed as \(z_{\text{A}, t}^{(i,j)} = p(\zeta_{i, j}|\boldsymbol{\mathrm{X}}_{\text{A}, t})\).

\subsection{Video tracker}
\label{subsec:video_tracker}
Video tracking is conducted using the \gls{YOLO} framework proposed in~\cite{Redmon2015}, which achieves a high detection rate in combination with real-time tracking capabilities. The implementation of YOLOFace is used, which is based on YOLOv3~\cite{Redmon2018}. Before an input video frame \(\boldsymbol{\mathrm{X}}_{\text{V}, t}\) is processed by the video tracker, it is first resized to \(832 \times 832\) pixels to meet the requirements of YOLOv3. We adapt the bounding box output to the \(20 \times 24\) grid size. The grid size can be set to arbitrary values if one is willing to retrain the network. We evaluated the localization performance using several grid sizes. However, we did not find any significant differences in localization performance. We decided to set the size of the individual grid-cells to roughly match the size of a human face in the dataset.
The video tracker's output probability is denoted as \(z_{\text{V}, t}^{(i,j)} = p(\zeta_{i, j}|\boldsymbol{\mathrm{X}}_{\text{V}, t})\).
\begin{figure}[t]
	\centering
	\includegraphics[width=0.65\linewidth]{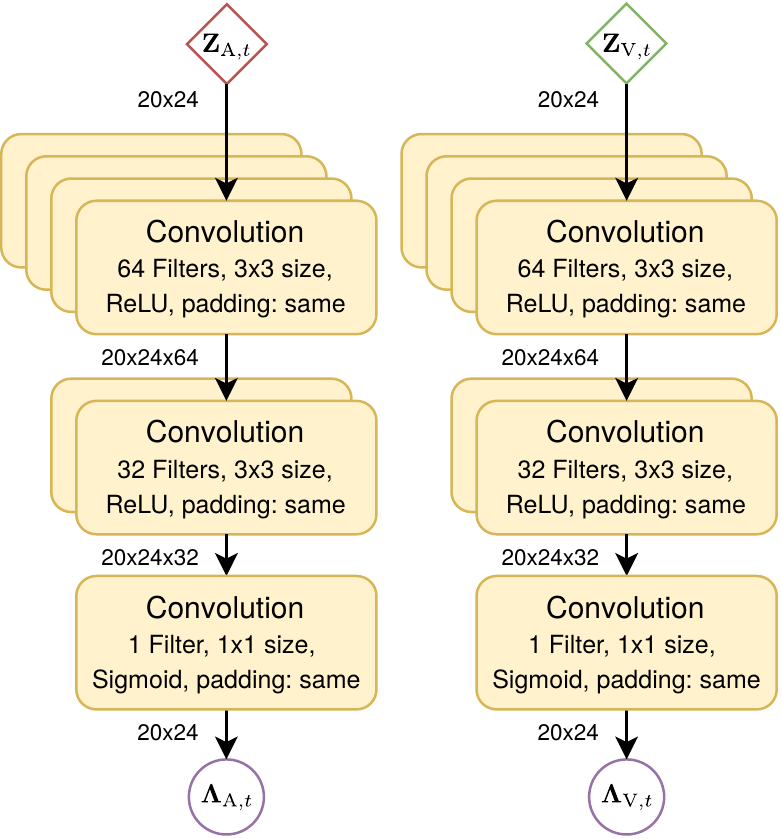}
		\vspace*{-0.1cm}
	\caption{Architecture of the acoustic and visual \gls{DSW} estimation networks, taking as 
	input the observations \(\boldsymbol{\mathrm{Z}}_{\text{A}, t}\) and \(\boldsymbol{\mathrm{Z}}_{\text{V}, t}\) obtained from the acoustic and visual trackers and producing as output the corresponding \glspl{DSW} \(\boldsymbol{\mathrm{\Lambda}}_{\text{A}, t}\) and \(\boldsymbol{\mathrm{\Lambda}}_{\text{V}, t}\) covering the  spatial grid.}
		\vspace*{-0.2cm}
	\label{fig:weight_network}
\end{figure}

\subsection{Data fusion}
\label{subsec:data_fusion}
At each time $t$, a conventional data fusion strategy would handle all acoustic and visual tracker outputs equally. In the log-domain, this \emph{flat fusion} strategy can be expressed as
\begin{equation}
    y_{t}^{(i,j)} = \log z_{\text{A}, t}^{(i,j)} + \log z_{\text{V}, t}^{(i,j)},
    \label{eqn:naive_fusion}
\end{equation}
where \(y_{t}^{(i,j)}\) represents the log-likelihood of a speaker being present in grid-cell \((i,j)\) at time-step \(t\), taking into account both acoustic and visual observations. Note that the derivation of Eq.~\eqref{eqn:naive_fusion} is based on assuming statistical independence between acoustic and visual observations, as well as imposing an uninformative prior on \(\zeta_{i, j}~\forall~i,j\), cf.~\cite{Schymura2019}.
However, when working with audiovisual input data, modalities may not be equally informative in all spatial regions due to, e.g., challenging lighting conditions or a directional noise source. In these cases, it should be possible to weight each source individually to lessen or increase the impact of an input. This process must be dynamic, in the sense that the network should be able to adapt over space and time. A \emph{weighted fusion} approach
\begin{equation}
    y_{t}^{(i,j)} = \lambda_{\text{A}, t}^{(i, j)} \log z_{\text{A}, t}^{(i,j)} + \lambda_{\text{V}, t}^{(i, j)}  \log z_{\text{V}, t}^{(i,j)}
    \label{eqn:weighted_fusion}
\end{equation}
is proposed here, where \(\lambda_{\text{A}, t}^{(i, j)}\), \(\lambda_{\text{V}, t}^{(i, j)} \in [0,\,1]\) denote the location- and time-dependent \glspl{DSW} for the acoustic and visual modalities, respectively.
In this manner, we are proposing an extension of the \gls{DSW} idea, originally introduced to handle only time-variant sensor reliability, to the spatial domain, allowing us to learn how different locations are covered more or less reliably by the available modalities. All relevant variables can be expressed in matrix notation as
\begin{equation*}
    \boldsymbol{\mathrm{Y}}_{t} = \begin{bmatrix}y_{t}^{(i,j)}\end{bmatrix}_{i,j},\, \boldsymbol{\mathrm{Z}}_{m, t} = \begin{bmatrix}z_{m, t}^{(i,j)}\end{bmatrix}_{i,j},\, \boldsymbol{\mathrm{\Lambda}}_{m, t} = \begin{bmatrix}\lambda_{m, t}^{(i,j)}\end{bmatrix}_{i,j},
\end{equation*}
with \(m\in \{\text{A}, \text{V}\}\), each covering the entire \(20 \times 24\) grid. The actual data fusion is performed conjointly with estimating the \glspl{DSW} in Eq.~\eqref{eqn:weighted_fusion} and the subsequent estimation of the speaker positions.
\subsection{Weight estimation}
\label{subsec:weight_estimation}

The individual acoustic and visual tracking networks discussed in Secs.~\ref{subsec:audio_tracker} and~\ref{subsec:video_tracker}, serve as inputs to the fusion neural network. The architecture of the corresponding \gls{DSW} estimation networks is depicted in Fig.~\ref{fig:weight_network}. Its output is subsequently combined with the outputs from both unimodal trackers according to Eq.~\eqref{eqn:weighted_fusion}.
%
To further refine the resulting speaker presence probability grid, it is fed into a dedicated refinement network based on the image restoration architecture proposed in~\cite{mao2016}. The parameterization of the model~\cite{mao2016} is adapted to the specific requirements of this work. In particular, the input and output size is set to \(20 \times 24\), matching the size of the probability grid. Each convolutional layer consists of 128 filters with a kernel size of \(3 \times 3\) and padding. The refinement operation is valuable to remove outliers and noisy estimates from the output grid, which should only contain peaks in grid cells close to the true speaker positions.

\section{Experimental setup}
In this section, we define the experimental settings and metrics\footnote{A Python implementation of all algorithms described in this paper is publicly available at https://github.com/rub-ksv/spatial-stream-weights}. 

\label{sec:evaluation}
\begin{figure*}[t]
\centering
\begin{minipage}[t][][b]{0.34\textwidth}\medskip
	\centering
	\includegraphics[width=0.99\linewidth]{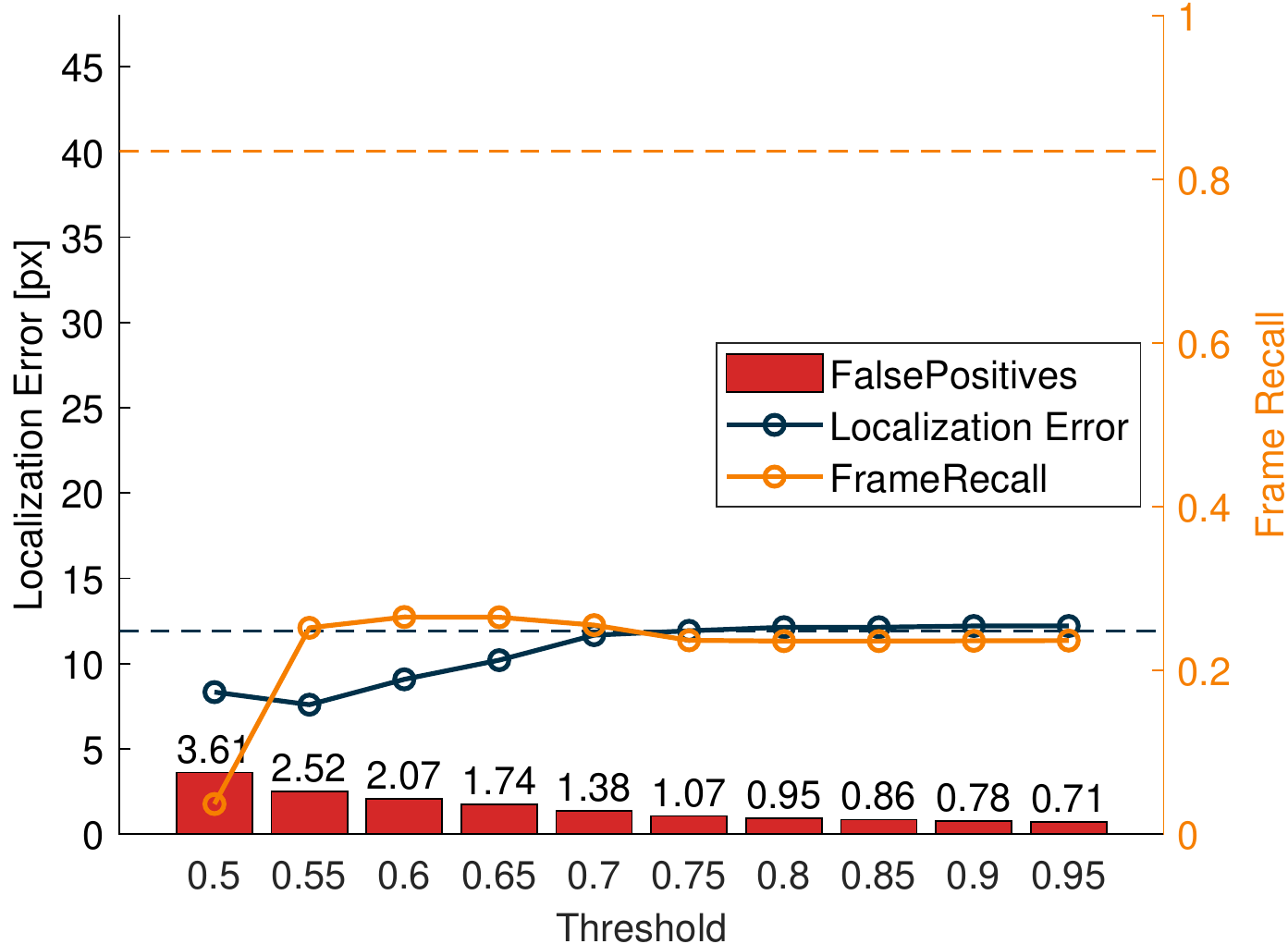}
	\label{fig:audio_metrics}
    \centerline{(a) Audio Baseline}\medskip
\end{minipage}%
\begin{minipage}[t][][b]{0.34\textwidth}\medskip
	\centering
	\includegraphics[width=0.99\linewidth]{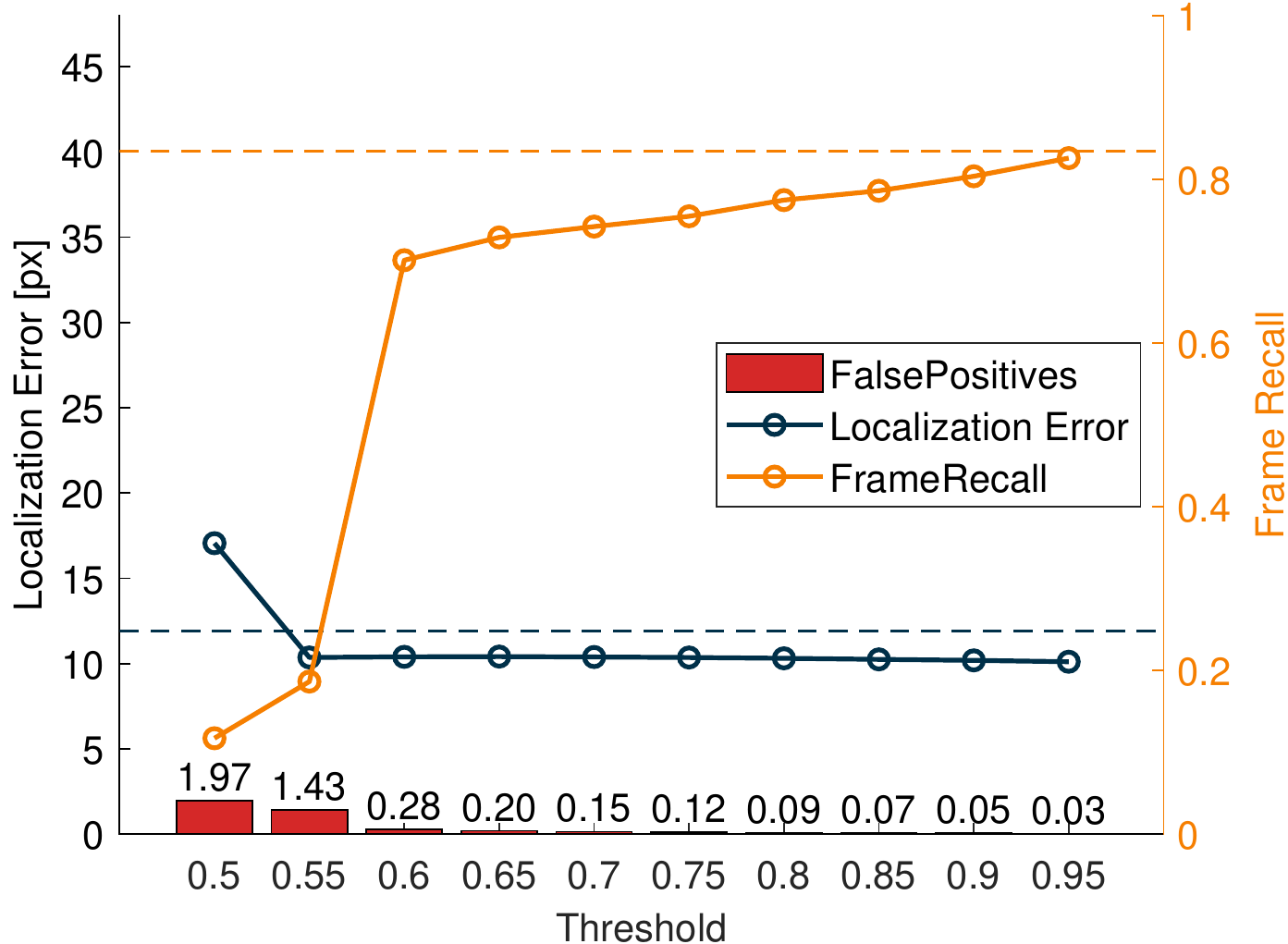}
	\label{fig:stream_fusion_time_variant_only}
    \centerline{(b) Spatially invariant \glspl{DSW}}\medskip
\end{minipage}%
\begin{minipage}[t][][b]{0.34\textwidth}\medskip
	\centering
	\includegraphics[width=0.99\linewidth]{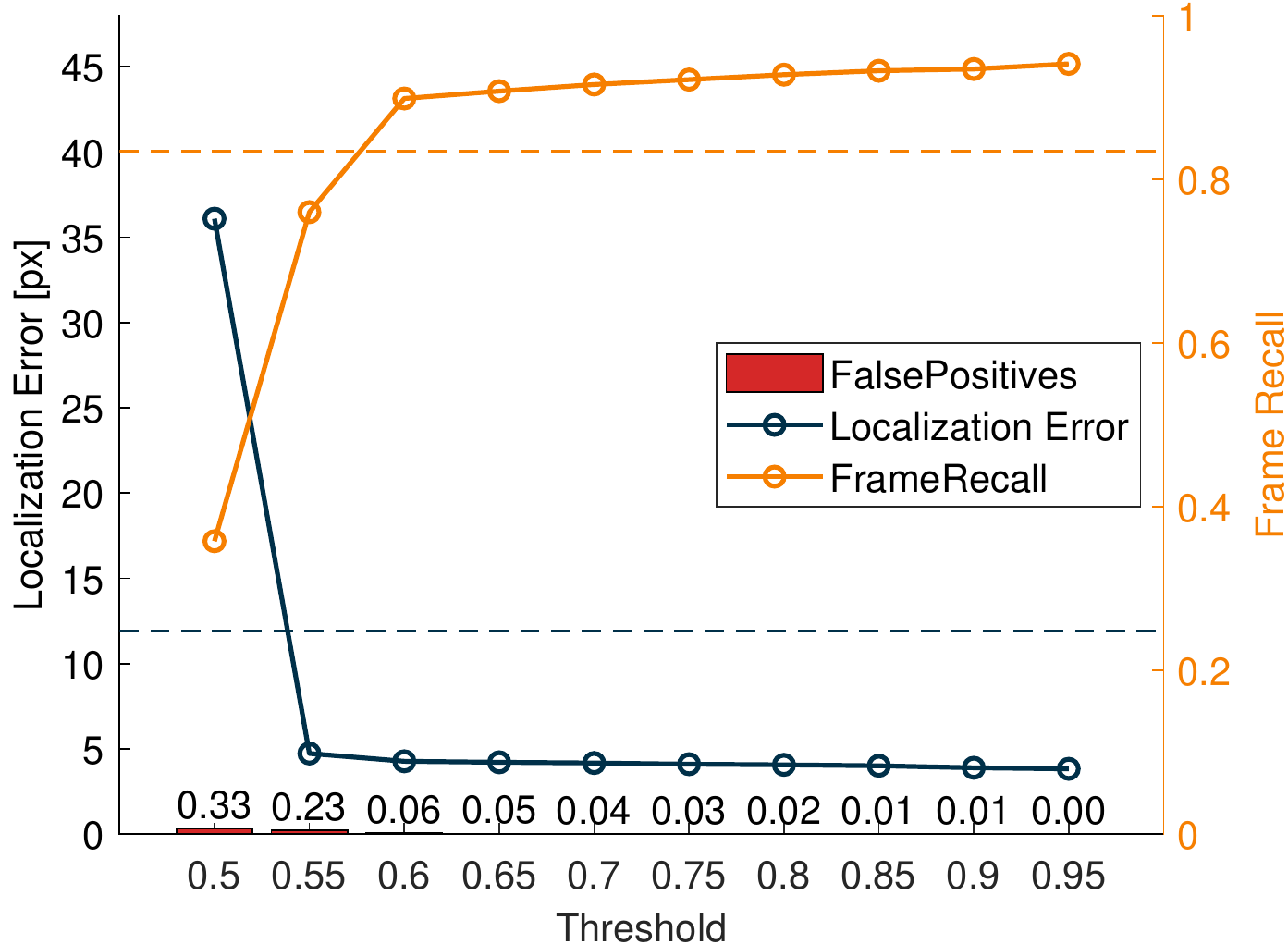}
	\label{fig:stream_fusion_weight_matrix_metrics}
    \centerline{(c) Weighted Fusion}\medskip
\end{minipage}%
\vspace*{-0.3cm}
\caption{Benchmark results plotted with respect to the threshold. The dashed lines represent the best unimodal baseline results.}
\vspace*{-0.2cm}
\end{figure*}

\subsection{Dataset}
\label{subsec:dataset}
The AVDIAR~\cite{Gebru2018} dataset is employed to evaluate the tracking capabilities of the proposed data fusion strategy. It contains 27 audiovisual sequences, which were recorded in three different rooms with a duration between 10 seconds and 3 minutes. The overall duration of the dataset is about 27 minutes. The recordings were conducted using 6 microphones attached to a dummy head and two cameras located right below the head. The microphone signals were acquired using a sampling rate of 48 kHz. The cameras cover a field of view of \(97\degree \times 80\degree\) at 25 frames per second. The video signals have a resolution of \(720 \times 450\) pixels. In this work, only one video signal from the left camera is used. The dataset was partitioned according to a training, validation and test split utilizing approximately \(75\%\), \(10\%\) and \(15\%\) of the available data, respectively. The splitting procedure ensured that all subsets included audiovisual sequences with one, two, three and four simultaneously visible speakers.

\subsection{Implementation details}
\label{subsec:experimental_setup}
All trainable weights were initialized using the Glorot initialization scheme~\cite{Glorot2010}. In our experiments, the Adam optimizer~\cite{DBLP:journals/corr/KingmaB14} with a batch size of 64, a learning rate of 0.001 and early stopping yielded the best results. The fusion networks were trained for 20 epochs, with early stopping triggering in the first five epochs with a patience of 4. The audio tracking networks were trained for 100 epochs, with early stopping triggering after 30 epochs with a patience of 20.

\subsection{Evaluation metrics}
\label{subsec:evaluation_metrics}
\Gls{LE} and \gls{FR}, as proposed in~\cite{Adavanne2018, Adavanne2019}, were utilized as evaluation metrics in this work. The \gls{LE} metric was slightly modified to cope with the discrete grid structure used here. Therefore, the Euclidean distance between a ground-truth speaker position and the center point of a grid cell where a speaker was detected (both represented as pixel coordinates), was utilized. This distance was computed for all possible combinations of detected speakers and ground-truth speaker positions at each time-frame. If fewer speakers were predicted than are actually present, the prediction vector was filled with the next highest activation, to have at least one prediction per target. To calculate the localization error for an estimated speaker position given the corresponding ground-truth, a cost matrix was optimized using the Hungarian algorithm~\cite{Kuhn1955}. The resulting distances were summed up and divided by the total number of speakers present at the particular time-frame, yielding the average distance per speaker. This process is repeated and averaged for the complete test set to obtain the final \gls{LE} metric. 


The \gls{FR} represents the ability of the network to predict the correct number of speakers in a specific frame. If the estimated number of speakers matches the ground-truth in that frame, the metric is set to one, otherwise it is zero. In the proposed framework, the number of speakers is set to the number of posterior grid probabilities surpassing the specified detection threshold, which has to be set empirically or via, e.g., a grid-search approach. Potentially, it might be included into an end-to-end optimization pipeline. Therefore, in Sec.~\ref{subsec:baseline}, different threshold values are evaluated for this purpose.

The \gls{FR} cannot differentiate between missed and excessive numbers of speakers. Therefore, we additionally count the average number of \glspl{FP} for each frame. This metric can be used to better understand and analyze causes of any detection issues. A low \gls{FR} in combination with a high \gls{FP} suggests that the network is not performing well due to the threshold being too low. On the other hand, in combination with a low \gls{FP} it seems reasonable that either the threshold is too high, pruning too many activations, or that the detector is not working properly.

\section{Results and discussion}
\label{sec:results_and_discussion}

In the following, we present and analyze the results of our baseline systems and the proposed fusion network architecture.

\subsection{Baseline methods}
\label{subsec:baseline}
The results of our two single-modality baselines, i.e.~audio- and video-only tracking systems, are provided in the first two rows of Tab.~\ref{tab:video}. As expected, the video-only baseline is generally superior in good visual conditions. The audio-only system outperforms video localization when faces are occluded or outside the field-of-view. 

With the single-modality audio tracker and a threshold of 0.75, an \gls{LE} of 11.93 pixels was achieved, while the \gls{FR} was quite low at 23\%. On average, a frame contained 1.07 false positives.
Fig.~4~(a) shows the performance of the audio tracker as a function of the threshold.
The \gls{LE} shows that the audio tracker predicts speakers with an error of roughly half a grid point. However, the low \gls{FR} indicates that many false positives are included in the tracking result. Even with a threshold of 0.75, on average, one false positive is included in every frame. 

\begin{table}[t]
\centering
\begin{tabular}{l||l|c|c|c|}
            & \gls{LE} [px]     & \gls{FR}      & \gls{FP}  \\ \hline \hline
Audio-only~(Sec.~\ref{subsec:audio_tracker})       & 11.93             & 0.24          & 0.072     \\ \hline
Video-only~(Sec.~\ref{subsec:video_tracker})       & 19.95             & 0.83          & 0.006     \\ \hline
Flat fusion~(Eq.~\ref{eqn:naive_fusion})     & ~~7.65              & 0.82          & 0.003     \\ \hline
\end{tabular}
\caption[Results of the baseline trackers.]{Results of the baseline trackers. For audio only the results for a threshold of 0.75 are given. For audio-visual tracking, we show the results of spatially flat fusion at a threshold of 0.75.}
\label{tab:video}
\end{table}

In comparison to the audio-only system, the video tracker in the second row in Tab.~\ref{tab:video} unsurprisingly achieves a significant higher \gls{FR} of 83.45\%, due to its access to a highly informative modality, in line with its good benchmark results in other works \cite{Redmon2018,Chen2020}. Note that no threshold is needed for the video tracker, as its output is binary. 
Additionally, the number of false positives per frame is lower, at 0.006. The \Gls{LE} is still below the size of one grid point with 19.95 pixels distance on average.

The face tracking capabilities lead to a good result with YOLO detecting the right number of faces in most cases. This is also reflected by the number of false positives, which are close to zero. 

Finally, the third row shows results of our naive audio-visual baseline system with the flat fusion, implemented according to Eq.~\eqref{eqn:naive_fusion}. As can be seen, the performance improves already, giving a lower localization error and false positive rate than the best uni-modal system at a virtually identical frame recall to the better-performing video setup.


\subsection{Spatially invariant fusion}
\label{subseq:spinvfu}
Fig.~4 (b) shows the results for the spatially invariant dynamic stream weighting approach. In this case, the DSWs in Eq.~\eqref{eqn:weighted_fusion} are simplified by using a time-variant but spatially invariant scalar, i.e.: $y_{t}^{(i,j)} = \lambda_{t}~ \log z_{\text{A}, t}^{(i,j)} + (1 - \lambda_{t})  ~\log z_{\text{V}, t}^{(i,j)}$. 

The dashed lines in Fig.~4 (b) represent the best reported single-modality baseline results for the \Gls{LE} and \Gls{FR}, respectively. 
The spatially invariant fusion is able to achieve a similar, but slightly worse \gls{FR} of 0.82. Taking the threshold into consideration, the \gls{FR} rises from 0.1 to above 0.7 for thresholds of 0.5 to 0.6. The reason is that the low activations around 0.5 of the audio-only tracker are carried over to the spatially invariant fusion and the flat fusion as well. Beyond a threshold of 0.6, the \gls{FR} increases slowly but steadily.

Comparing the \glspl{LE} in Fig.~4 (b) with the reported values in Tab.~\ref{tab:video}, we can observe that incorporating additional information to the data fusion layer leads to further improvements in the model's ability to extrapolate the ground-truth location of the speakers: With temporal information, we see an improvement from 11.9 to 10.1 pixels. With the flat fusion strategy reported in Tab.~\ref{tab:video}, we would see a lower LE, reduced from 10.1 to 7.7 pixels.

Through multi-modal fusion, the number of false positives also improves w.r.t.~the respective best-performing unimodal system, which is again similarly observable for the flat fusion strategy. Note that lower thresholds generally return worse results and choosing a higher threshold helps to mask out outliers. Increasing the threshold beyond 0.55 does not lead to any further changes in the LEs and FPs, indicating that outliers are sufficiently suppressed at this threshold.

\subsection{Spatially weighted fusion}
Fig.~4 (c) shows the performance of the weighted fusion system. 
At a threshold of 0.75, the network achieves a \gls{FR} of 0.92, clearly outperforming all the other strategies. Again, the low performance at a threshold of 0.5 mentioned in Sec.~\ref{subseq:spinvfu} is still visible. 

In terms of the LE, the best result was achieved with 4.10 pixels distance at a threshold of 0.75. For all thresholds above 0.5, nearly constant results were measured. The FPs per frame were relatively low, at an average of 0.028 over-predictions per frame. 

Overall, the spatially weighted fusion strategy with a location-dependent matrix achieved the best performance. The target speakers were located with a high precision, showing the lowest \gls{LE} measured across all experiments. Even though the number of false positives was increased slightly, the overall capabilities of the network are significantly better compared to the other systems. 

Finally, comparing Fig.~4~(b) and~(c), the questions arises, whether the improved performance may be caused by the ability to focus on different modalities in different parts of the room and whether this ability may be helpful in dealing with multiple speakers. To investigate these questions, we compared the performance metrics for audio frames containing only a single speaker with audio frames containing $4$ different speakers. 
As hypothesized, we achieved the same results (LE = 1.31, FR = 1 and FP = 0) for both fusion strategies, when only a single speaker is involved. In contrast, when 4 speakers are present, the performances differ significantly. While the performance of the spatially weighted fusion strategy remains stable (LE = 0,53, FR = 0,99, FP = 0), the LE of the spatially invariant fusion strategy suffers (LE = 10,4, FR = 0,81, FP = 0). 
While based on comparatively little data, this analysis appears to indicate that spatial weighting is better equipped for dealing with multi-speaker scenarios.

\section{Conclusion}
\label{sec:conclusion_and_outlook}

In this paper, we introduced a novel deep-learning-based data-fusion strategy by extending the idea of dynamic temporal stream weights to the spatial domain. Experiments show that the unimodal baseline video tracker based on the YOLO-framework achieved accurate results as a stand-alone system, while, unsurprisingly, using only the unimodal audio tracker led to a significantly higher false positive rate. However, by fusing acoustic and visual information, it was possible to achieve an improved tracking result, with a 10\% increase of the frame recall and a far more accurate localization, showing an error of 3.83 pixels on average. The results show that the audiovisual tracking system benefits from the proposed data fusion strategy, and that adding spatial dependence of stream weights is highly beneficial compared to a purely temporal dependence, as in standard dynamic stream weighting. 





\vfill\pagebreak

\bibliographystyle{IEEEbib}
\bibliography{refs}

\begin{thebibliography}{10}

\bibitem{Mendes2016}
J.~J.~A. Mendes, M.~Vieira, M.~B. Pires, and S.~L. Stevan,
\newblock ``Sensor fusion and smart sensor in sports and biomedical
  applications,''
\newblock {\em Sensors (Basel, Switzerland)}, vol. 16, no. 10, September 2016.

\bibitem{Kam1997}
M.~{Kam}, X.~{Zhu}, and P.~{Kalata},
\newblock ``Sensor fusion for mobile robot navigation,''
\newblock {\em Proceedings of the IEEE}, vol. 85, no. 1, pp. 108--119, 1997.

\bibitem{Li2015}
W.~Li, Z.~Wang, G.~Wei, L.~Ma, J.~Hu, and D.~Ding,
\newblock ``A survey on multisensor fusion and consensus filtering for sensor
  networks,''
\newblock {\em Discrete Dynamics in Nature and Society}, vol. 2015, pp. 683701,
  Oct 2015.

\bibitem{Hershey2000}
J.~R. Hershey and J.~R. Movellan,
\newblock ``Audio {Vision}: Using audio-visual synchrony to locate sounds,''
\newblock in {\em Advances in Neural Information Processing Systems 12}, S.~A.
  Solla, T.~K. Leen, and K.~M\"{u}ller, Eds., pp. 813--819. MIT Press, 2000.

\bibitem{Ban2018}
Y.~Ban, X.~Alameda{-}Pineda, Laurent Girin, and R.~Horaud,
\newblock ``Variational {Bayesian} inference for audio-visual tracking of
  multiple speakers,''
\newblock {\em CoRR}, vol. abs/1809.10961, 2018.

\bibitem{Alameda-Pineda2019}
X.~Alameda-Pineda, S.~Arias, Y.~Ban, G.~Delorme, L.~Girin, R.~Horaud, X.~Li,
  B.~Morgue, and G.~Sarrazin,
\newblock ``Audio-visual variational fusion for multi-person tracking with
  robots,''
\newblock in {\em Proceedings of the 27th ACM International Conference on
  Multimedia}, New York, NY, USA, 2019, MM '19, p. 1059–1061, Association for
  Computing Machinery.

\bibitem{Gebru2014}
I.~D. {Gebru}, X.~{Alameda-Pineda}, R.~{Horaud}, and F.~{Forbes},
\newblock ``Audio-visual speaker localization via weighted clustering,''
\newblock in {\em 2014 IEEE International Workshop on Machine Learning for
  Signal Processing (MLSP)}, Sep. 2014, pp. 1--6.

\bibitem{GaticaPerez2003}
D.~{Gatica-Perez}, G.~{Lathoud}, I.~{McCowan}, J.~. {Odobez}, and D.~{Moore},
\newblock ``Audio-visual speaker tracking with importance particle filters,''
\newblock in {\em Proceedings 2003 International Conference on Image
  Processing}, 2003, vol.~3, pp. III--25.

\bibitem{GaticaPerez2007}
D.~{Gatica-Perez}, G.~{Lathoud}, J.~{Odobez}, and I.~{McCowan},
\newblock ``Audiovisual probabilistic tracking of multiple speakers in
  meetings,''
\newblock {\em IEEE Transactions on Audio, Speech, and Language Processing},
  vol. 15, no. 2, pp. 601--616, Feb 2007.

\bibitem{Nakadai2002}
K.~{Nakadai}, K.~{Hidai}, H.~G. {Okuno}, and H.~{Kitano},
\newblock ``Real-time speaker localization and speech separation by
  audio-visual integration,''
\newblock in {\em Proceedings 2002 IEEE International Conference on Robotics
  and Automation}, 2002, vol.~1, pp. 1043--1049.

\bibitem{Busso2005}
C.~{Busso}, S.~{Hernanz}, C.-W. {Chu}, S.-I. {Kwon}, S.~{Lee}, P.~G.
  {Georgiou}, I.~{Cohen}, and S.~{Narayanan},
\newblock ``Smart room: participant and speaker localization and
  identification,''
\newblock in {\em Proceedings. (ICASSP '05). IEEE International Conference on
  Acoustics, Speech, and Signal Processing, 2005.}, 2005, vol.~2, pp.
  ii/1117--ii/1120 Vol. 2.

\bibitem{Gebru2018}
I.~D. Gebru, S.~Ba, X.~Li, and R.~Horaud,
\newblock ``{Audio-Visual Speaker Diarization Based on Spatiotemporal Bayesian
  Fusion},''
\newblock {\em {IEEE Transactions on Pattern Analysis and Machine
  Intelligence}}, vol. 40, no. 5, 2018.

\bibitem{kalman1960}
R.~E. Kalman,
\newblock ``A new approach to linear filtering and prediction problems,''
\newblock {\em Journal of Basic Engineering}, vol. 82, no. 1, pp. 35--45, 03
  1960.

\bibitem{Meutzner2017}
H.~{Meutzner}, N.~{Ma}, R.~{Nickel}, C.~{Schymura}, and D.~{Kolossa},
\newblock ``Improving audio-visual speech recognition using deep neural
  networks with dynamic stream reliability estimates,''
\newblock in {\em 2017 IEEE International Conference on Acoustics, Speech and
  Signal Processing (ICASSP)}, March 2017, pp. 5320--5324.

\bibitem{Schoenherr2016}
L.~Sch\"onherr, D.~Orth, M.~Heckmann, and D.~Kolossa,
\newblock ``{Environmentally robust audio-visual speaker identification},''
\newblock in {\em 2016 IEEE Spoken Language Technology Workshop (SLT)}, Dec
  2016.

\bibitem{Schymura2019}
C.~{Schymura} and D.~{Kolossa},
\newblock ``Audiovisual speaker tracking using nonlinear dynamical systems with
  dynamic stream weights,''
\newblock {\em IEEE/ACM Transactions on Audio, Speech, and Language
  Processing}, vol. 28, pp. 1065--1078, 2020.

\bibitem{Schymura2020}
C.~{Schymura}, T.~{Ochiai}, M.~{Delcroix}, K.~{Kinoshita}, T.~{Nakatani},
  S.~{Araki}, and D.~{Kolossa},
\newblock ``A {Dynamic Stream Weight} backprop {Kalman} filter for audiovisual
  speaker tracking,''
\newblock in {\em ICASSP 2020 - 2020 IEEE International Conference on
  Acoustics, Speech and Signal Processing (ICASSP)}, 2020, pp. 581--585.

\bibitem{Adavanne2018}
S.~{Adavanne}, A.~{Politis}, and T.~{Virtanen},
\newblock ``Direction of arrival estimation for multiple sound sources using
  convolutional recurrent neural network,''
\newblock in {\em 2018 26th European Signal Processing Conference (EUSIPCO)},
  Sep. 2018, pp. 1462--1466.

\bibitem{Redmon2015}
J.~Redmon, S.~K. Divvala, R.~B. Girshick, and A.~Farhadi,
\newblock ``You only look once: Unified, real-time object detection,''
\newblock {\em CoRR}, vol. abs/1506.02640, 2015.

\bibitem{Redmon2018}
J.~Redmon and A.~Farhadi,
\newblock ``{YOLOv3}: An incremental improvement,''
\newblock {\em CoRR}, vol. abs/1804.02767, 2018.

\bibitem{mao2016}
X.~Mao, C.~Shen, and Y.~Yang,
\newblock ``Image restoration using very deep convolutional encoder-decoder
  networks with symmetric skip connections,''
\newblock in {\em Advances in Neural Information Processing Systems 29}, D.~D.
  Lee, M.~Sugiyama, U.~V. Luxburg, I.~Guyon, and R.~Garnett, Eds., pp.
  2802--2810. Curran Associates, Inc., 2016.

\bibitem{Glorot2010}
X.~Glorot and Y.~Bengio,
\newblock ``Understanding the difficulty of training deep feedforward neural
  networks,''
\newblock in {\em Proceedings of the Thirteenth International Conference on
  Artificial Intelligence and Statistics}, Yee~Whye Teh and Mike Titterington,
  Eds., Chia Laguna Resort, Sardinia, Italy, 13--15 May 2010, vol.~9 of {\em
  Proceedings of Machine Learning Research}, pp. 249--256, PMLR.

\bibitem{DBLP:journals/corr/KingmaB14}
D.~P. Kingma and J.~Ba,
\newblock ``Adam: {A} method for stochastic optimization,''
\newblock in {\em 3rd International Conference on Learning Representations,
  {ICLR} 2015, San Diego, CA, USA, May 7-9, 2015, Conference Track
  Proceedings}, Y.~Bengio and Y.~LeCun, Eds., 2015.

\bibitem{Adavanne2019}
S.~{Adavanne}, A.~{Politis}, J.~{Nikunen}, and T.~{Virtanen},
\newblock ``Sound event localization and detection of overlapping sources using
  convolutional recurrent neural networks,''
\newblock {\em IEEE Journal of Selected Topics in Signal Processing}, vol. 13,
  no. 1, pp. 34--48, 2019.

\bibitem{Kuhn1955}
H.~W. Kuhn,
\newblock ``The {Hungarian} method for the assignment problem,''
\newblock {\em Naval Research Logistics Quarterly}, vol. 2, no. 1‐2, pp.
  83--97, 1955.

\bibitem{Chen2020}
W.~Chen, H.~Huang, S.~Peng, C.~Changsheng, and C.~Zhang,
\newblock ``{YOLO}-{Face}: a real-time face detector,''
\newblock {\em The Visual Computer}, Mar 2020.

\end{thebibliography}

\end{document}